\documentclass[12pt,manuscript]{aastex}

\begin{document}

\shorttitle{OSIRIS Mask Designer}

\shortauthors{Gonz\'alez-Serrano et al.}

 \title {\bf OSIRIS software: the Mask Designer Tool}

\author{J.I. Gonz\'alez-Serrano}
\affil{Instituto de F\'\i sica de Cantabria (CSIC-Universidad de
Cantabria)} \affil{Avda. de los Castros s/n, 39005 Santander,
Spain} \email{gserrano@ifca.unican.es}
\author{M. S\'anchez-Portal}
\affil{Universidad Pontif\'\i cia de Salamanca en Madrid, 28040
Madrid, Spain}
\author{H. Casta\~neda}
\affil{Instituto de Astrof\'\i sica de Canarias, 38200 La Laguna,
Tenerife, Spain  }
\author{R. Quirk, E.D. de Miguel}
\affil{GMV S.A., 28760 Tres Cantos, Madrid, Spain} \and
\author{M. Aguiar, J. Cepa}
\affil{Instituto de Astrof\'\i sica de Canarias, 38200 La Laguna,
Tenerife, Spain}

\begin{abstract}

OSIRIS is a Day One instrument that will be available at the 10m
GTC telescope which is being built at La Palma observatory in the
Canary Islands. This optical instrument is designed to obtain
wide-field narrow-band images using tunable filters and to do
low-resolution spectroscopy in both long-slit and multislit modes.
For the multislit spectroscopy mode, we have developed a software
to assist the observers to design focal plane masks. In this paper
we describe the characteristics of this Mask Designer tool. We
discuss the main design concepts, the functionality and particular
features of the software.

\end{abstract}

\section{INTRODUCTION}

OSIRIS (Optical System for Imaging and low/intermediate-Resolution
Integrated Spectroscopy) is an instrument designed to obtain images
and low-resolution spectra in the optical domain (from 3650 through
10,000 \AA). It will be installed at the Nasmyth focus of the 10m
GTC telescope  \citep{alva04} at La Palma observatory, although it
can be also mounted at the Cassegrain focus. A detailed account of
the observing modes, optical design, and performance is given in
\citet{cepa03}.

OSIRIS is a versatile instrument offering many observing modes: (a)
direct imaging with broad-band and narrow-band  using tunable
filters, (b) narrow-band with charge shuffling, (c) fast photometry,
(d) long-slit spectroscopy, (e) multi-object spectroscopy, (f)
nod-and-shuffle multi-object spectroscopy, and (g) fast
spectroscopy. The main driver of the optical and mechanical design
of OSIRIS has been the inclusion of tunable filters (TF) in order to
obtain narrow-band imaging at an arbitrary wavelength with effective
resolutions of around 5-60 \AA. In the spectroscopic modes, OSIRIS
provides spectral resolutions $R=\lambda/\Delta\lambda$ of 250, 500,
1500, and 2500 covering the full spectral range from 3650 to 10,000
\AA\ by using 11 grisms. Grisms can be rotated by $90^\circ$ to
change the dispersion direction and allow for fast spectroscopic
observations. The detector consists of a mosaic of two 2k$\times$4k
Marconi MAT-44-82 CCDs of 15$\mu$m/pixel providing a large field of
view ($\sim 8\farcm5\times8\farcm5$) and a pixel size of
$0\farcs125$. There will be a gap of 8.3 arcsec parallel to the
normal dispersion direction.

OSIRIS will be the first instrument using the technique of tunable
filters in large telescopes. The potential of this observational
mode has been described widely in the literature (e.g.
\citet{jones01}; \citet{jones02}; \citet{bland03};
\citet{glaz04}).

The next observational mode in complexity is the multi-object
spectroscopy (MOS). This technique arrived at its maturity in the
eighties, increasing significantly the number of objects that can be
observed simultaneously. Together with advances in optical design,
new coatings for the camera elements, large-format detectors, and
new technologies in grisms and diffraction gratings, their use in
large telescopes is opening new windows in astronomical research,
especially in the developments of large scale surveys. Recently, the
use of the technique of nod-and-shuffle allows to obtain spectra
with optimal sky subtraction, making possible to observe deeper
\citep{glaz01}.

Most of the observing modes need to block part of the FOV to allow
either charge shuffling or slit and multislit spectroscopy. This
blocking is done by using a mask located at the telescope focal
plane. Masks are fabricated by cutting the necessary holes or
slits in aluminium plates which are of $496\times 480$ mm in size,
being the size of the imaging field of $422\times 428$ mm. In
standard MOS mode, the number of slits (8 arcsec long) that can
fit in a single mask is around 60 and 100 at the highest and
lowest spectral resolution, respectively.

 In this paper we describe the software that has been
developed to assist the observer to design  OSIRIS masks.

\section{OVERVIEW OF THE OSIRIS MASK SYSTEM}

The OSIRIS Mask System is the system in charge of storing and
positioning masks and it is fully described in \citet{mili03}.
Figure 1 shows a general view of the Mask System. A cassette
provides storage for 13 masks which are positioned at the focal
plane by a Mask Loader Mechanism. The position accuracy of the
masks over the telescope focal plane is of the order of 0.01 mm.

Masks are slightly curved in cylindrical form in order to adapt to
the geometry of the focal plane. Masks will be tooled in a flat
plate and afterwards bent to fit the cylindrical shape over the
mask frame.

\section{OVERVIEW OF THE MASK DESIGNER}

For each 10m-class telescope where an instrument with MOS
capabilities is operating, a specific software for mask design has
been developed. Examples are the FORS Instrument Mask Simulator
\citep{humm00}, the VIMOS Mask Preparation Software
\citep{botti01}, and the Mask Design Program of FOCAS
\citep{saito03}. Our software has been designed to provide similar
capabilities to these programs, adapted to the special needs of
OSIRIS.

The Mask Designer is a subsystem of the OSIRIS Observing Program
Manager which is, in turn, integrated into the GTC Control System
\citep{filg98}. The software has been designed using Unified
Modelling Language (UML) and coded in Java programming language.

The OSIRIS Mask Designer is a plug-in application of the
JSkyCat\footnote{\tt http://archive.eso.org/JSky} package. In this
way, the software uses classes from JSkyCat to manage FITS files,
to perform coordinate transformations, and for graphical user
interface.

 The Mask Designer performs all the
calculations needed to transform sky (RA and DEC) and CCD ($x,y$)
coordinates to coordinates over the mask. The final output of the
Mask Designer is therefore a set of instructions for the machine in
charge of cutting the holes in the mask plate. This means that
accurate coordinate transformations are needed to convert from sky
to telescope focal plane positions, and from CCD to focal plane
positions. Also, atmospheric refraction and thermal expansion
corrections should be taken into account. Figure 2 shows all the
coordinate transformations needed to correct a sky position onto a
CCD position which include: (1) atmospheric refraction, (2) gnomonic
projection and geometrical distortion of the telescope focal plane,
(3) transformation from telescope focal plane onto mask surface, (4)
mapping from mask to CCD plane. Atmospheric refraction is computed
following \citet{fili82} and depends on wavelength, air pressure and
temperature, as well as on zenith distance of the targets.
Transformations between different surfaces are made using 5th-order
polynomials whose coefficients will be obtained in the instrument
commissioning phase. Taking into account the errors due to mask
fabrication, mechanical flexures, thermal expansion, astrometric
accuracy, and the fact that the mask shape does not match exactly
the focal plane, we expect that the position of the slits over the
targets (after acquiring the field using fiducial stars), will be
better than a 6\% of the slit width. For a nominal slit of 1.4
arcsec, this corresponds to a positional accuracy of 0.08 arcsec.

Two types of masks can be generated depending of the OSIRIS
observing mode. In the standard MOS mode, slits can be located at
any position over the physical area of the mask. The second
observing mode is nod and shuffle which allows astronomers to
subtract the sky emission by nodding the telescope between targets
and sky while simultaneously CCD charge is shuffled up and down
(e.g. \citet{glaz01}). In this mode slits can be smaller than in
standard MOS mode.

The system performs the necessary checks for mask integrity, slit
collision and spectra overlapping. In case of overlapping slits or
spectra the software can assign slits to additional masks using a
priority-based algorithm. The software also takes into account the
gap in the detector plane and the edges of the mask plate which
are disabled by the mask frame that holds the mask plate.

 \section{DECOMPOSITION DIAGRAM}

Figure 3 is a diagram showing the high-level classes in which the
Mask Designer system has been decomposed. These classes
(UserInputHandler, FileHander, and MaskHandler) are in turn
decomposed into individual classes.
 The class OSIRIS\_Mask\_Designer represents the
whole system and contains all the functionality to fully produce
masks. Composite class UserInputHandler holds the functionality
for handling the user interaction with the system. FileHander
class is in charge of all actions regarding files. The MaskHandler
composite class contains the algorithms for processing the slit
data, transforming coordinates, and producing valid mask
configurations. Its decomposition diagram is shown in Fig. 3.

Algorithms for processing the slits and producing valid mask
configurations belong to classes SlitProcessor and MaskProcessor.
Classes SkyToMaskMapping and CCD\_ToMaskMapping contain the
functions needed to transform coordinates between the different
systems involved.

Not all the operations are shown in this diagram as all classes
are decomposed into many simpler functions.

The Mask Designer has been designed as a plug-in of JSkyCat
application, which contains many other classes and packages
required by the system. The most relevant JSky utilities and
classes used by the OSIRIS Mask Designer are: WCSTransform to
convert between image pixel and WCS (World Coordinate System)
coordinates; FITSCodec for handling of FITS files; GUI classes for
image display and related tasks.

\section{FEATURES OF THE MASK DESIGNER}

The OSIRIS Mask Designer is launched by starting the JSkyCat
application and therefore, its functionality initially is
identical to that of JSkyCat. All operations are invoked through
the main JSkyCat menu bar (see Fig. 4).

\subsection{Configuration file}

For the different parameters related to a particular observation,
such as date, hour angle, central coordinates of the field, grism,
position of the grism (0 or 90 degrees for normal or fast modes,
respectively), position angle of the mask onto the sky, expected
observation temperature and pressure, expected fabrication
temperature, maximum number of masks per pointing, the system
implements a Configure Observation menu to read an XML file
(Observation Configuration File). Other XML configuration files
are needed by the system but they are not user configurable. These
contain telescope and instrument parameters and information
relevant to computations, such as the coefficients for coordinate
transformations and grism properties. The main advantage of using
XML files is that full independence from target platform is
guaranteed.

\subsection{Input options}

The software can process slits defined interactively by using an
image displayed on the screen, via cursor. This image should
either have absolute astrometry information on its header or it
can be a raw frame taken with OSIRIS in broad-band mode.
Alternatively, the user can provide a list of coordinates and slit
properties (type, width, priority) that can be, in turn, either
absolute equatorial coordinates or pixel coordinates from a raw
broad-band image taken with OSIRIS. These options allow processing
object lists obtained with, for instance, image detection
software.

The mask designer can be run in four different modes:

\begin{itemize}

\item Defining slit positions interactively using a raw image
taken with OSIRIS

\item Defining slit positions interactively using an image
containing absolute astrometric information in its header

\item Using a list of RA and DEC coordinates

\item Using a list of $x$, $y$ positions from a raw image
taken with OSIRIS

\end{itemize}

\subsection{Slits}

Three different types of slits can be defined: (a) rectangular,
(b) circular, and (c) curved slits. Circular slits are used mainly
(but not exclusively) to define fiducial holes. A minimum of three
fiducial slits must be defined, normally to include bright stars
in the field, in order to acquire the field. The type of the slit
is defined by launching the Slit Window (see Fig. 5) and can be
changed at any moment during the process of defining the slits.
Also, the slit dimensions can be changed by selecting a slit
already defined. Slits can be deleted either one by one or all at
once.

A special feature of this software is the ability of defining
curved slits (available only in interactive processing). These are
arc-shaped slits and may be useful for taking spectra of special
targets such as, for instance, galaxy arcs produced by
gravitational lensing.

\subsection{Overlapping and integrity check}

Integrity checks are made to avoid placing slits too close to mask
borders or to other slits which may cause the mask to break at the
fabrication process. Once all the slits are successfully
configured the system processes the information. Several masks can
be generated up to a maximum value specified by the user. The
system will try to place as many slits as possible in the first
mask. Slits are placed in additional masks depending on their
priority, which is set by the user. Overlapping slits or slits
producing integrity conflicts are placed onto the second mask and
so forth until the maximum number of masks is reached. A residual
mask is produced containing conflicting or out-of-bounds slits.
This mask will not be sent to fabrication, being its only purpose
to inform the user.

Conflicting and out-of-bounds slits are highlighted on the display
using a color code to help the user to re-design the mask.

\subsection{Output}

The output consists of several files containing all the
information required to build the masks.  The user can save the
work after masks have been processed. Saved files can be retrieved
for later processing if needed. There are three output files, two
of them in FITS format and the third in a format that will be read
by the cutting machine software. A FITS file called ODF (Object
Definition File) includes all the slits defined by the user,
including invalid ones. The file contains the observation and
telescope data, and, for all the generated masks, slit type and
slit positions. A separate storage is provided in the ODF file to
keep the fiducial slit information. This file can be handled by
the Mask Designer to allow the user to modify the design in new
sessions.

 A second FITS file, called MDF (Mask Definition File),
is generated for each mask. There is one file per mask containing
only the slit information corresponding to that mask. A MDF can be
saved also for the Residual Mask, although in that case, fiducial
slits information is lost.
 The third file is the so-called Mask Cutter File and contains
 instructions to be read by the cutting machine to fabricate
 physically the masks.

The user has the option of printing a graphic view of each mask,
containing the defined slits, the fiducial holes and, if wanted,
rectangles marking where the spectra will appear in the image. An
example is shown in Figure 6.

\section{CONCLUSIONS}

We have designed and developed an application to design focal
plane multislit masks for the instrument OSIRIS. Its functionality
has been tested and the user requirements have been achieved.
 The software takes into account all the effects
that could affect an accurate slit positioning over the mask
plates, and allows different input scenarios to design OSIRIS
masks. It has been designed as a JSky plug-in offering an easy
 user interface.

 Only during the instrument commissioning phase will be
possible to test the software in real conditions. A critical issue
is the fine tuning of the parameters involved in coordinate
transformations. In particular, the coefficients to convert from
telescope focal plane to detector and from sky to telescope focal
plane will be obtained using pin-hole masks and  observations of
astrometric fields. Commisioning phase for OSIRIS at the GTC is
planned to start in 2006.

\acknowledgments Figure 1 has been kindly provided by Lorenzo
Peraza. This work has been supported by the Spanish Plan Nacional
de Astronom\'\i a y Astrof\'\i sica under grants AYA2000-2688-E,
AYA2002-03326, and AYA2002-01379.

\figcaption[]{General view of the OSIRIS Mask System showing the
focal plane bench, the mask frame, and the mask cassette with
capacity for 13 masks}

\figcaption[]{Schematic view of the coordinate transformations
involved in the design of a slit mask}

 \figcaption[]{OSIRIS Mask Designer decomposition diagram showing, at the highest level, the classes in which
 the Mask Designer is decomposed. A further class decomposition is shown here for class MaskHandler, which is in charge
 of the most important functions of the application}

\figcaption[]{Mask Designer Menu selection in JSkyCat}

\figcaption[]{Slit Window used to define and change slit
properties}

\figcaption[]{Example of output showing the slits and the
predicted spectra over the detector area of OSIRIS. There are 68
slits including four fiducial circular holes and a curved slit.
The horizontal band at the middle of the plot represents the gap
between the two CCD detectors}

\newpage

Keywords: telescopes -- instrumentation: spectrographs

\newpage


\begin{thebibliography}{}

\bibitem[\'Alvarez and Rodr\'\i guez-Espinosa
(2004)]{alva04}{\'Alvarez, P. and Rodr\'\i guez-Espinosa, J.M.
2004, Proceedings of the SPIE, vol. 5489, 583}

\bibitem[Bland-Hawthorn and Kedziora-Chudczer
(2003)]{bland03}{Bland-Hawthorn, J. and Kedziora-Chudczer, L.
2003, PASA, 20, 242}

\bibitem[Bottini, Garilli, and Tresse (2001)]{botti01}{Bottini,
D., Garilli, B, Tresse, L. 2001,  in ASP Conf. Ser. 238,
Astronomical Data Analysis Software and Systems X, ed. F.R.
Harnden, Jr., F.A. Primini, \& H.E. Payne (San Francisco: ASP)
455}

\bibitem[Cepa et al. (2003)]{cepa03}{Cepa, J. et al. 2003,
Proceedings of the SPIE, vol. 4841, 1739}

\bibitem[Filgueira and Rodr\'\i guez (1998)]{filg98}{Filgueira,
J.M. and Rodr\'\i guez, D. 1998, Proceedings of the SPIE, vol.
3351, 2}
\bibitem[Filippenko (1982)]{fili82}{Filippenko, A.V. 1982, PASP,
94, 715}
\bibitem[Glazebrook and Bland-Hawthorn (2001)]{glaz01}{Glazebrook,
K. and Bland-Hawthorn, J. 2001, PASP, 113, 197}

\bibitem[Glazebrook, et al. (2004)]{glaz04}{Glazebrook, K.,
Tobber, J., Thomson, S., Bland-Hawthorn, J., Abraham, R. 2004, AJ,
128, 2752}

\bibitem[Hummel (2000)]{humm00}{Hummel, W. 2000, Proceedings of
the SPIE, vol 4010, 190}

\bibitem[Jones and Bland-Hawthorn (2001)]{jones01}{Jones, D.H. and
Bland-Hawthorn, J. 2001, ApJ, 550, 593}

\bibitem[Jones, Shopbell, and Bland-Hawthorn (2002)]{jones02}{Jones,
D.H., Shopbell, P.L., Bland-Hawthorn, J. 2002, MNRAS, 329, 759}

\bibitem[Militello et al. (2003)]{mili03}{Militello, C., Correa,
S., P\'erez, J., Fuentes, F.J., P\'erez de Taoro, R., Cepa, J.,
Peraza, L. 2004, Proceedings of the SPIE, vol. 4841, 1515}

\bibitem[Saito et al. (2003)]{saito03}{Saito, Y., et al. 2003,
Proceedings of the SPIE, vol. 4841, 1180}

\end{thebibliography}
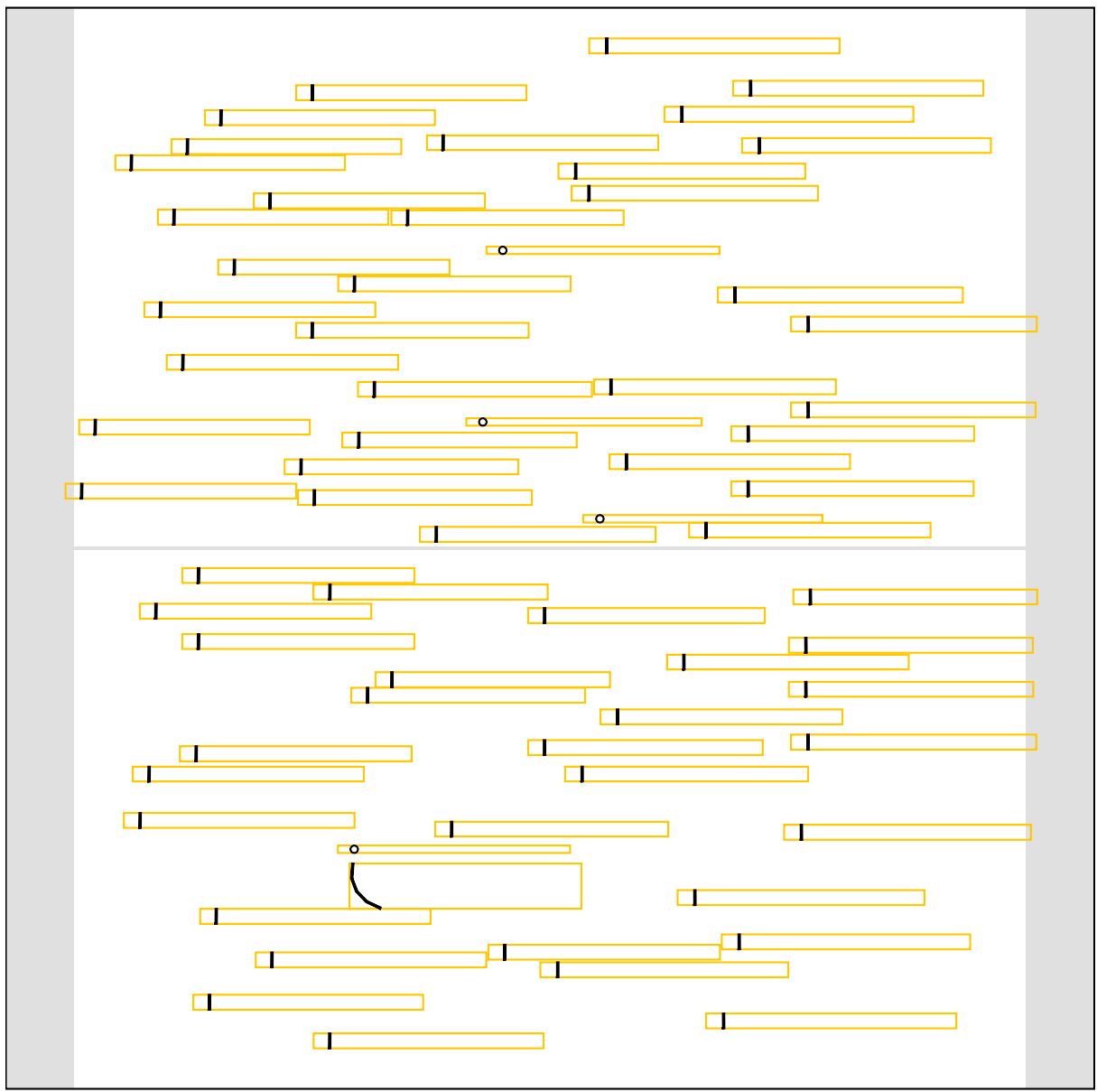
\end{document}